\pdfoutput=1
\documentclass[amsmath,amssymb,prl,
showpacs,superscriptaddress,preprintnumbers]{revtex4}
\usepackage{graphicx}
\usepackage{latexsym}
\usepackage{bm}
\usepackage{color}

\begin{document}

\title{$\theta_{13}$, $\mu\tau$ symmetry breaking and neutrino Yukawa textures}

%\preprint{SINP/APC/2010-\ldots}

\author{Biswajit Adhikary}
\email[Electronic mail: ]{biswajit.adhikary@saha.ac.in}
\affiliation{Saha Institute of Nuclear Physics, 1/AF Bidhannagar, Kolkata 700064, India}

\affiliation{Department of Physics, Gurudas College, Narkeldanga,
Kolkata 700054, India}

\author{Ambar Ghosal}
\email[Electronic mail: ]{ambar.ghosal@saha.ac.in}
\affiliation{Saha Institute of Nuclear Physics, 1/AF Bidhannagar, Kolkata 700064, India}

\author{Probir Roy}
\email[Electronic mail: ]{probir.roy@saha.ac.in}
\affiliation{Saha Institute of Nuclear Physics, 1/AF Bidhannagar, Kolkata 700064, India}
\begin{abstract}
\noindent
Within the type-I seesaw and in the basis where charged lepton and heavy neutrino mass matrices are real and
 diagonal, $\mu\tau$ symmetric four and three zero neutrino Yukawa textures are perturbed by lowest order $\mu\tau$
symmetry breaking terms. These perturbations are taken to be the {\it most general ones for those textures}.
 For quite small values of those symmetry breaking parameters, 
permitting a lowest order analysis, current best-fit ranges of 
neutrino mass squared 
differences and mixing angles are shown to be accommodable, including a value of $\theta_{13}$ in the observed range, 
provided all the light neutrinos have an inverted mass ordering. 
%\end{abstract}
\end{abstract}
\maketitle
%%%%%%%%%%%%%%%%%%%%%%%%%
%%%%%%%%%%%%%%%%%%%%%%%%%%%%%%%%%%%%%%%%%%%%%%%%%%%%%%%%%%%%%
\section{Introduction}
A major recent development in Particle Physics has been the 
observation \cite{Fpan,Ahn:2012nd,Abe:2011fz,Abe:2011sj} of a 
significant mixing 
between the first and third generations of (anti-) neutrinos with a measured angle $\theta_{13}=8.8^\circ\pm1.1^\circ$.
 The underlying  physical implication is rather serious. Certain flavor symmetries in the neutrino sector, such as 
 that under $\mu\leftrightarrow\tau$ interchange \cite{Grimus:2012hu} (i.e. $2\leftrightarrow 3$
 in relevant matrix elements), which imply a vanishing $\theta_{13}$, must be broken. 
The latter became a highly popular idea on account
of its prediction of a maximal mixing ($\theta_{23}=45^\circ$) 
between the second and third generations of 
neutrinos --- a situation still well-allowed by extant data. 
But, now that $\theta_{13}$ is sizably nonzero, 
there is interest in breaking this $\mu\tau$ symmetry. Its spontaneous
breakdown generally requires 
\cite{Grimus:2012hu,Joshipura:2009tg} 
several additional fields. So explicit and radiative modes of $\mu\tau$ symmetry breaking 
are the only viable options with the choice of a minimal set of fields. 
However, 
radiative breaking, related to a high scale scenario with 
$\theta_{13}=0$ through Renormalisation 
Group Evolution,
is characterised by \cite{Dighe:2008wn} a small loop-induced constant 
proportional to $(m_\tau/v)^2$, $v$ being the EW VEV. The latter, 
even in the 
light neutrino mass ordering case producing the largest $\theta_{13}$, 
is unable \cite{Dighe:2008wn,Dighe:2007ksa}  to generate a $\theta_{13}$  
 larger than $5^\circ$ which is somewhat
disfavoured by the data. A reasonably minimalist approach then
 would be to try the explicit breaking of $\mu\tau$ symmetry as a perturbation.

 We have in  mind a canonical type-I seesaw 
\cite{Min,Gell,Yana,Mohapatra:1979ia} mechanism for the generation of the light neutrino Majorana
 mass matrix $M_\nu$ with three heavy ($>10^9$ GeV) 
right chiral EW singlet Majorana neutrinos. 
We prefer to introduce the 
perturbation in the neutrino Yukawa coupling matrix or equivalently in 
the neutrino Dirac
 mass matrix $M_D$. The latter, rather than $M_\nu$, is what appears in the Lagrangian. Hence
the present approach
reflects a more basic way of handling the above explicit symmetry breaking. Our next step is to study the effects
of those parameters on {\it predictive} neutrino Yukawa textures, specially on four and three 
zero $\mu\tau$ symmetric Yukawa textures of $M_D$. Such Yukawa texture zeros may arise in a number of models, e.g. 
those with \cite{Harigaya:2012bw} extra dimensions. It may be recalled here that the study of presumed four zero 
 textures has had a distinguished record in the quark sector 
\cite{Fritzsch,Fritzsch:1999ee, Babu:2004tn}. It is thus natural to extend 
 similar ideas to neutrinos modulo the difference due to the type-I seesaw. 
One can raise the issue of arbitrariness in our choice of textures, but 
our preference for the maximal number of zeros in the $\mu\tau$ symmetric 
neutrino Yukawa coupling matrix is motivated by predictivity. Textures with 
fewer zeros have many more parameters and do not have testable predictions.
Any texture statement is, of course,  dependent 
 on the weak basis chosen. We select one in which the charged lepton and heavy right chiral neutrino 
mass matrices are real, positive and diagonal.

By an $n$ zero texture, what we mean here is an allowed configuration of $M_D$ with $n$ vanishing elements.
Three and four zero textures provide a predictive and useful framework \cite{Adhikary:2012zx} within which one can discuss neutrino 
masses and mixing angles. As already mentioned, this utility gets much 
reduced for texures with a fewer number of
zeros. Exact $\mu\tau$ symmetry automatically yields 
$\theta_{23}=\pi/4$ and $\theta_{13}=0$. Additionally, each of the allowed 
textures leads to an $M_\nu$ with three independent parameters (two real ones and one phase) apart from an overall mass
factor. Our aim here is to study deviations from these 
consequences of $\mu\tau$ symmetry due to the symmetry breaking
perturbation. 

In deciding which $\mu\tau$ symmetric textures are allowed and which are not, we are guided by twin criteria. First,
 the observed fact of none of the three light neutrinos being unmixed in flavor means that a block diagonal form of 
$M_D$ is inadmissible. Further, if any row of $M_D$ is orthogonal, element by element 
(ruling out unnatural cancellations) to each of the other two, one neutrino family 
decouples -- disallowing that 
texture. Second, in the absence of any fundamental principle dictating as such, none of the neutrinos is taken 
to be strictly massless, i.e ${\rm det}\,\, M_\nu\ne 0$, which 
requires via the seesaw that ${\rm det}\,\, M_D\ne 0$. The presence 
of three heavy right chiral singlet Majorana neutrinos is crucial here since,
with only two, a massless neutrino is inevitable \cite{tanimoto}. 
This means that no entire row or column of $M_D$ can vanish, nor can there be in it a quartet of zeros at the corners
 of a rectangular array.

With the above constraints, four was shown \cite{Branco:2007nb} to be the maximum number of zeros allowed in a texture of $M_D$. All
 such four zero textures have been discussed extensively 
\cite{Choubey:2008tb,Adhikary:2012kb,Liao:2013kix}. These textures were further restricted 
\cite{Adhikary:2009kz,Adhikary:2011pv} drastically to
 four allowed ones by the imposition of $\mu\tau$ symmetry. The four allowed 
textures, named $A1$, $A2$ and $B1$, $B2$ 
(each pair yielding the same $M_\nu$), were of course included in the 
complete list of Ref. \cite{Branco:2007nb}. They have 
also been shown to be capable of leading to a desirable level of baryogenesis via leptogenesis 
\cite{Adhikary:2010fa}. Interestingly,
 allowed three zero textures of $M_D$, when made $\mu\tau$ symmetric, are also found to be drastically reduced in 
number to only two. These are designated $C1$ and $C2$ here, both leading to the same $M_\nu$.

Each one of the $M_D$ textures $A1$, $A2$, $B1$, $B2$, $C1$, $C2$ is then 
perturbed away from $\mu\tau$ symmetry
by deviation factors of the form $1-\epsilon_ie^{i\phi_i}$ (no sum). 
We now call the perturbed texture $M_D^\epsilon$.
It should be emphasized that, while the perturbations explicitly break $\mu\tau$ symmetry, the maximal zero texture 
is kept intact. The latter is our basic framework which is retained without change.
Here the parameters $\epsilon_i$ are real positive numbers, kept small in order that higher order terms can be neglected. Further, the phases $\phi_i$ 
are unrestricted except for being between $-\pi$ and $+\pi$. Because of the presence of two independent off-diagonal elements in the
$\mu\tau$ symmetric form of any of $M_{DA1}$, $M_{DA2}$, $M_{DB1}$ and $M_{DB2}$, only two
 such deviation factors are needed per texture in complete generality.
 In contrast, the $\mu\tau$ symmetric form of either of 
$M_{DC1}$, $M_{DC2}$ has three independent off-diagonal elements; hence three deviation factors each have to be 
inserted in general in these cases. Additionally, a deviation factor $1-\delta$, $\delta$ being real, is introduced
 in the third element of the diagonal $\mu\tau$ symmetric form of the right chiral heavy neutrino mass matrix $M_R$.

The corresponding complex symmetric 
light neutrino Majorana mass matrix $M^{\epsilon,\delta}_\nu$ 
obtains via the type-I seesaw from the above 
perturbed texures. Thus we have 
\begin{equation}
 M^{\epsilon,\delta}_\nu\simeq -M_D^{\epsilon}(M^{\delta}_{R})^{-1}{M_D^{\epsilon}}^T.
\label{mnuepd}
\end{equation}
 Of course, one needs to carefully follow the interplay between the number of independent parameters in the emergent
$M_\nu$ and the number of separate experimental inputs, as was 
emphasized \cite{Merle:2006du} 
 some time ago. With the parametric
 form of  $M^{\epsilon,\delta}_\nu$ for each of the six textures, we construct
the hermitian product
\begin{equation}
 H^{\epsilon,\delta}_\nu= M^{\epsilon,\delta}_\nu{ M^{\epsilon,\delta}_\nu}^\dagger.
\label{hepd}
\end{equation}
Through $H^{\epsilon,\delta}_\nu$ we directly connect with 
five experimentally measured quantities of phenomenological relevance, to wit
$\Delta^2_{21}=m_2^2-m_1^2$, $|\Delta^2_{32}|=|m_3^2-m_2^2|$, 
$\theta_{12}$, $\theta_{23}$ and $\theta_{13}$.

The rest of the paper is organized as follows. In Section 2 
we outline our basic theoretical framework. Section 3
contains an enumeration of the forms of the light neutrino mass 
matrix $M_\nu$ from $\mu\tau$ symmetric allowed four zero 
and three zero textures. The corresponding perturbed and broken $\mu\tau$ symmetric expression for $M^{\epsilon,\delta}_\nu$ 
in parametric form is constucted for each case in Section 4 and the corresponding 
$H^{\epsilon,\delta}_\nu$ is displayed. Section 5 is devoted to a phenomenological discussion of 
what is allowed/disallowed in the space of parameters from the five experimental inputs and which textures are 
in/out, given the $3\sigma$ ranges of those measured quantities. Finally, Section 6 contains a summary of our 
conclusions. In the Appendix we provide some details of the diagonalization procedure.  
%%%%%%%%%%%%%%%%%%%%%%%%%%%%%%%%%%%%%%%%%%%%%%%%%%%%%%%%%%%%%%%%%
\section{Three and four zero textures with broken $\mu\tau$ symmetry}
%%%%%%%%%%%%%%%%%%%%%%%%%%%%%%%%%%%%%%%%%%%%%%%%%%%%%%%%%%%%%%%%%%
We give a compact outline of our theoretical framework here since a major 
part of it was detailed earlier 
\cite{Adhikary:2012zx,Adhikary:2009kz,Adhikary:2011pv}. We have already 
referred to the type-I seesaw and the H matrix in eqns.(1) and (2) 
in the Introduction.
These can be considered without the $\epsilon$ superscript in the unperturbed 
limit. We can write
%   
%The general type-I seesaw  formula for the light neutrino Majorana mass matrix $M_\nu$ can be written 
%in the usual notation as 
%\begin{eqnarray}
%M_\nu\simeq-M_DM_R^{-1}M_D^T
%\label{ss}
%\end{eqnarray}
%
%\begin{eqnarray}
%H_\nu=M_\nu M_\nu^\dagger
%\label{u1}
%\end{eqnarray}
\begin{eqnarray}
U^\dagger H^{\epsilon,\delta}_\nu U= {\rm diag}\,(|m_1|^2, |m_2|^2,|m_3|^2),
\label{u}
\end{eqnarray} 
\noindent
where we have followed the PDG convention \cite{Beringer:1900zz} in defining 
$m_1$, $m_2$,$m_3$.
In (\ref{u}), $U$ is the unitary PMNS matrix. 
Under $\mu\tau$ symmetry, $M_3 = M_2$, but - as mentioned earlier - 
the broken $\mu\tau$ symmetric extension 
%of (\ref{u1}) 
can be given as
\begin{eqnarray}
M_R={\rm diag}\,\left[M_1,~M_2,~M_2(1-\delta)\right],
\label{mtbmr}
\end{eqnarray}
where $\delta$ is a real parameter which can have either sign. 
In the $\mu\tau$ symmetric limit, 
$\delta\rightarrow0$ and moreover 
$M_{12}=M_{13}$, $M_{21}=M_{31}$, $M_{22}=M_{33}$ and $M_{23}=M_{32}$ where $M$ can be either $M_D$ or $M_\nu$.\\
\newpage
\underline{Four zero textures}\\

On applying the twin criteria explained  in the Introduction  and $\mu\tau$ symmetry, only four textures of $M_D$ are
found to survive. They are divided pairwise into two categories $\it{A}$ and $\it{B}$ and are individually called
$A1,~A2$ and $B1,~B2$. In terms of arbitrary complex quantities $a$, $b$, $c$, they can be written as
\begin{eqnarray}
M_{DA1}^{(4)}=\left(\begin{array}{ccc} a & b & b\\ 0 & 0 & c\\0 & c & 0 \end{array}
\right),\qquad
M_{DA2}^{(4)}=\left(\begin{array}{ccc} a & b & b\\ 0 & c & 0\\0 & 0 & c \end{array}
\right),\nonumber\\
 M_{DB1}^{(4)}=\left(\begin{array}{ccc} a & 0 & 0\\ b & 0 & c\\b & c & 0 \end{array}
\right),\qquad
M_{DB2}^{(4)}=\left(\begin{array}{ccc} a & 0 & 0\\ b & c & 0\\b & 0 & c \end{array}
\right).
\label{al40}
\end{eqnarray}
The last two are respective transposes of the first two and the superscripts '$4$' signifies a four zero texture. 
Each pair in a category yields the same $M_\nu^{(4)}$. With a reparametrisation in terms of an overall complex mass
factor $m_{A/B}$ and two real positive quantities $k_{1,2}$/$l_{1,2}$ and a phase $\alpha$/$\beta$ (cf. Table I),
they appear in Category $A$/$B$ as  
\begin{eqnarray}
M_{\nu A}^{(4)} &=& m_A\left(\begin{array}{ccc}k_1^2e^{2i\alpha}+2k_2^2&k_2&k_2\cr
                        k_2 &1& 0\cr
                        k_2&0&1\end{array}\right), \nonumber\\
M_{\nu B}^{(4)} &=& m_B \left(\begin{array}{ccc}
             l_1^2&l_1l_2e^{i\beta}&l_1l_2e^{i\beta}\cr
                                    l_1l_2e^{i\beta}&l_2^2e^{2i\beta}+
1&l_2^2e^{2i\beta}\cr
                                    l_1l_2e^{i\beta}&l_2^2e^{2i\beta}
&l_2^2e^{2i\beta}+1
\end{array}\right).
\label{mnu4030}
\end{eqnarray}
 \begin{table*}[ht]
 \caption{ \label{param} Parameters in $M_\nu$ and functions $X_{1,2,3}$. The unphysical primed phases can be rotated away.}
 \begin{tabular}{|c|c|c|c|}
 \hline
  Type of Textures&\multicolumn{2}{|c|}{Four zero} & Three zero\\
\cline{1-4}
Category&A& B& C\\
 \hline
% Category& A1~~~~~~~~~~~~A2 & B1~~~~~~~~~~~~~B2&C1~~~~~~~~~~~~~C2\\
% \hline
% Textures of $M_D$ & $\left(\begin{array}{ccc} a & b & b\\ 0 & 0 & c\\0 & c & 0 \end{array}
 %\right)$ $\left(\begin{array}{ccc} a & b & b\\ 0 & c & 0\\0 & 0 & c \end{array}
 %\right)$ & $\left(\begin{array}{ccc} a & 0 & 0\\ b & 0 & c\\b & c & 0 \end{array}
 %\right)$ $\left(\begin{array}{ccc} a & 0 & 0\\ b & c & 0\\b & 0 & c \end{array}
 %\right)$ & $\left(\begin{array}{ccc} 0 & b & b\\ c & 0 & d\\c & d & 0 \end{array}\right)$
 % $\left(\begin{array}{ccc} 0 & b & b\\ c & d & 0\\c & 0 & d \end{array}\right)$\\
 %\hline
 %$M_\nu$ & $m_A\left(\begin{array}{ccc}k_1^2e^{2i\alpha}+2k_2^2&k_2&k_2\cr
 %                        k_2 &1& 0\cr
  %                       k_2&0&1\end{array}\right)$& $m_B \left(\begin{array}{ccc}
  %          l_1^2&l_1l_2e^{i\beta}&l_1l_2e^{i\beta}\cr
 %                                    l_1l_2e^{i\beta}&l_2^2e^{2i\beta}+
 %1&l_2^2e^{2i\beta}\cr
  %                                   l_1l_2e^{i\beta}&l_2^2e^{2i\beta}
 %&l_2^2e^{2i\beta}+1
 %\end{array}\right)$ & $m_C\left(\begin{array}{ccc} 2r_1^2  &  r_1  & r_1 \\ 
 %r_1  & r_2^2e^{2i\gamma}+1 & r_2^2e^{2i\gamma}\\
 %r_1  & r_2^2e^{2i\gamma} &  r_2^2e^{2i\gamma}+1 \end{array}\right)$\\
 %\hline
 Definition&$m_A = -c^2/M_2,$  & $m_B = -c^2/M_2, $ & $m_C=-d^2/M_2$\\
of &&&\\
\cline{2-4}
 parameters&$k_1e^{i(\alpha+\alpha^\prime)}=\frac{a}{c}\sqrt{\frac{M_2}{M_1}},$ & 
 $l_1 e^{i\beta^\prime}=\frac{a}{c}\sqrt{\frac{M_2}{M_1}}$& ${r_1 e^{i\gamma^\prime}} = \frac{b}{d}$\\
\cline{2-4}
 in $M_\nu$&$k_2 e^{i\alpha^\prime}= \frac{b}{c},$  & $l_2e^{i\beta} = \frac{b}{c}\sqrt{\frac{M_2}{M_1}}$& 
 $r_2e^{i\gamma}=\frac{c}{d}{\sqrt \frac{M_2}{M_1}}$ \\
\cline{2-4}
 &$\alpha = {\rm arg}\frac{a}{b}$ & $\beta = {\rm arg}\frac{b}{c}$ & $\gamma = {\rm arg}\frac{c}{d}$\\
 \hline 
 $X_1$&$2\sqrt{2}k_2[{(1+2k_2^2)}^2 + k_1^4 $ & $2\sqrt{2}l_1l_2[{(l_1^2+2l_2^2)}^2 + 
 1$ & $2{\sqrt 2}r_1[(1+2r_1^2)^2+4r_2^4$\\
 & $+ 2k_1^2(1+2k_2^2)\cos2\alpha]^{1/2}$ &$+ 2(l_1^2+2l_2^2)\cos2\beta]^{1/2}$ &$+4r_2^2(1+2r_1^2)\cos2\gamma]^{1/2}$\\
 \hline
 $X_2$& $1-k_1^4-4k_2^4-4k_1^2k_2^2\cos2\alpha
 $ & $ 1+4l_2^2\cos2\beta+4l_2^4-l_1^4
 $ & $4r_2^4+1+4r_2^2\cos2\gamma -4r_1^4$\\
 \hline
$X_3$&$1-4k_2^4-k_1^4 -4k_1^2k_2^2\cos2\alpha $ & $1-{(l_1^2+2l_2^2)}^2$
& $1-4r_1^4-4r_1^2 -4r_2^4$\\
&$-4k_2^2$ & $-4l_2^2\cos2\beta$ & $-4r_2^2\cos2\gamma$\\
\hline
\end{tabular} 
\end{table*}
The most general $\mu\tau$ symmetry breaking perturbation on $M_D$ consits of two independent complex terms 
 containing $\epsilon_1e^{i\phi_1}$, $\epsilon_2e^{i\phi_2}$. The four 
textures of (\ref{al40}) are then extended to
\begin{eqnarray}
M_{DA1}^{\epsilon(4)}&=&\left(\begin{array}{ccc}a & b & b(1-\epsilon_1e^{i\phi_1})\\ 
             0 & 0 & c(1-\epsilon_2e^{i\phi_2})\\
             0 & c  & 0\end{array}\right),~~~~~~~~~~~~~~\qquad%\nonumber\\
 M_{DA2}^{\epsilon(4)}=\left(\begin{array}{ccc}a & b & b(1-\epsilon_1e^{i\phi_1})\\ 
             0 & c(1-\epsilon_2e^{i\phi_2}) & 0\\
             0 & 0  & c\end{array}\right),\nonumber\\
M_{DB1}^{\epsilon(4)}&=&\left(\begin{array}{ccc}a & 0 & 0\\ 
             b(1-\epsilon_1e^{i\phi_1}) & 0 & c(1-\epsilon_2e^{i\phi_2})\\
             b & c  & 0\end{array}\right),\qquad%\nonumber\\ 
M_{DB2}^{\epsilon(4)}=\left(\begin{array}{ccc}a & 0 & 0\\ 
             b(1-\epsilon_1e^{i\phi_1}) & c(1-\epsilon_2e^{i\phi_2}) & 0\\
             b & 0  & c\end{array}\right).\nonumber\\
 \label{mdmtb40}
 \end{eqnarray}
We always work to the lowest order in the epsilons and $\delta$. The seesaw 
enables the reparametrisation of
$M^{\epsilon,\delta(4)}_\nu$ in terms of $m_{A/B}$, $k_{1,2}$/$l_{1,2}$, $\alpha$/$\beta$ while including $\mu\tau$
 symmetry breaking terms involving $\epsilon_1e^{i\phi_1}$, $\epsilon_2e^{i\phi_2}$ and $\delta$. Whereas, in the
 $\mu\tau$ symmetric limit, there are only two $M_\nu^{(4)}$'s, now there are four $M_\nu^{\epsilon,\delta(4)}$'s:
\begin{eqnarray}
M_{\nu A1}^{\epsilon,\delta(4)} &=& m_A\left(\begin{array}{ccc}k_1^2e^{2i\alpha}+
2k_2^2(1-\epsilon_1e^{i\phi_1}+\frac{\delta}{2})&k_2(1-\epsilon_1e^{i\phi_1}-\epsilon_2e^{i\phi_2}+\delta)&k_2\cr
                        k_2(1-\epsilon_1e^{i\phi_1}-\epsilon_2e^{i\phi_2}+\delta) &1-2\epsilon_2e^{i\phi_2}
+\delta& 0\cr
                        k_2&0&1\end{array}\right), \nonumber\\\nonumber\\
M_{\nu A2}^{\epsilon,\delta(4)} &=& m_A\left(\begin{array}{ccc}k_1^2e^{2i\alpha}+
2k_2^2(1-\epsilon_1e^{i\phi_1}+\frac{\delta}{2})&k_2(1-\epsilon_2e^{i\phi_2})&
k_2(1-\epsilon_1e^{i\phi_1}+\delta)\cr
                        k_2(1-\epsilon_2e^{i\phi_2}) &1-2\epsilon_2e^{i\phi_2} & 0\cr
                        k_2(1-\epsilon_1e^{i\phi_1}+\delta)&0&1+\delta\end{array}\right), \nonumber\\\nonumber\\
M_{\nu B1}^{\epsilon,\delta(4)} &=& m_B \left(\begin{array}{ccc}
             l_1^2&l_1l_2e^{i\beta}(1-\epsilon_1e^{i\phi_1})&l_1l_2e^{i\beta}\cr
 l_1l_2e^{i\beta}(1-\epsilon_1e^{i\phi_1})&l_2^2e^{2i\beta}(1-2\epsilon_1e^{i\phi_1}-2\epsilon_2e^{i\phi_2})
+1+\delta &l_2^2e^{2i\beta}(1-\epsilon_1e^{i\phi_1})\cr
                                    l_1l_2e^{i\beta}&l_2^2e^{2i\beta}(1-\epsilon_1e^{i\phi_1})
&l_2^2e^{2i\beta}+1
\end{array}\right),\nonumber\\\nonumber\\
M_{\nu B1}^{\epsilon,\delta(4)} &=& m_B \left(\begin{array}{ccc}
             l_1^2&l_1l_2e^{i\beta}(1-\epsilon_1e^{i\phi_1})&l_1l_2e^{i\beta}\cr
 l_1l_2e^{i\beta}(1-\epsilon_1e^{i\phi_1})&l_2^2e^{2i\beta}(1-2\epsilon_1e^{i\phi_1}-2\epsilon_2e^{i\phi_2})
+1 &l_2^2e^{2i\beta}(1-\epsilon_1e^{i\phi_1})\cr
                                    l_1l_2e^{i\beta}&l_2^2e^{2i\beta}(1-\epsilon_1e^{i\phi_1})
&l_2^2e^{2i\beta}+1+\delta
\end{array}\right).\nonumber\\
\label{mtbmnu40}
\end{eqnarray}
\underline{Three zero textures}\\

Our twin criteria here leave only two $\mu\tau$ symmetric textures as survivors, each with a vanishing ($1,1$) 
element. We categorise them under the general 
designation Category $C$, calling them $C1$ and $C2$. Thus
\begin{eqnarray}
M_{DC1}^{(3)}&=&\left(\begin{array}{ccc} 0 & b & b\\ c & 0 & d\\c & d & 0 \end{array}\right), \nonumber\\ 
 M_{DC2}^{(3)}&=&\left(\begin{array}{ccc} 0 & b & b\\ c & d & 0\\c & 0 & d \end{array}\right),
\label{al30}
\end{eqnarray}
with $b$, $c$, $d$ being arbitrary complex quantities in general. 
In analogy with four zero textures the single seesaw induced mass matrix
$M_{\nu C}^{(3)}$ in this case can be reparametrised (cf Table I) 
in terms of real positive quantities $r_{1,2}$, a 
phase $\gamma$ and an overall complex mass factor 
$m_C$ as   
\begin{eqnarray}
 M_{\nu C}^{(3)}&=&m_C\left(\begin{array}{ccc} 2r_1^2  &  r_1  & r_1 \\ 
 r_1  & r_2^2e^{2i\gamma}+1 & r_2^2e^{2i\gamma}\\
 r_1  & r_2^2e^{2i\gamma} &  r_2^2e^{2i\gamma}+1 \end{array}\right).
\label{mnu30}
\end{eqnarray}
Here too the superscript $(3)$ refers to the three zero texture origin.

For these textures there can be three independent $\mu\tau$ symmetry breaking perturbing terms in general. We can therefore extend the textures of
(\ref{al30}) to
\begin{eqnarray}
M_{DC1}^{\epsilon(3)}=\left(\begin{array}{ccc}0 & b & b(1-\epsilon_1e^{i\phi_1})\\ 
              c(1-\epsilon_2e^{i\phi_2}) & 0 & d(1-\epsilon_3e^{i\phi_3})\\
              c & d  & 0\end{array}\right),%\nonumber\\
  M_{DC2}^{\epsilon(3)}=\left(\begin{array}{ccc}0 & b & b(1-\epsilon_1e^{i\phi_1})\\ 
              c(1-\epsilon_2e^{i\phi_2}) & d(1-\epsilon_3e^{i\phi_3}) & 0\\
              c & 0  & d\end{array}\right).\nonumber\\
 \label{mdmtb30}
 \end{eqnarray}

Working to the lowest order in $\epsilon_{1,2,3}$, 
the neutrino mass matrix of (\ref{mnu30}) splits into two in terms of 
the reparametrised quantities $r_{1,2},~\gamma$ and $m_C$ as follows:
\begin{eqnarray}
 M_{\nu C1}^{\epsilon,\delta(3)}&=&m_C\left(\begin{array}{ccc} 2r_1^2(1-\epsilon_{1}e^{i\phi_1}+\frac{\delta}{2})  &  r_1(1-\epsilon_{3}e^{i\phi_3})  & r_1(1+\delta) \\ 
 r_1(1-\epsilon_{3}e^{i\phi_3})  & r_2^2e^{2i\gamma}(1+2\epsilon_{2}e^{i\phi_2})  +1-2\epsilon_{3}e^{i\phi_3} & r_2^2e^{2i\gamma}(1-\epsilon_{2}e^{i\phi_2})  \\
 r_1(1+\delta)  & r_2^2e^{2i\gamma}(1-\epsilon_{2}e^{i\phi_2}) &  r_2^2e^{2i\gamma}+1+\delta \end{array}\right),\nonumber\\
 M_{\nu C2}^{\epsilon,\delta(3)}&=&m_C\left(\begin{array}{ccc} 2r_1^2(1-\epsilon_{1}e^{i\phi_1}+\frac{\delta}{2})  & 
 r_1(1-\epsilon_{1}e^{i\phi_1}-\epsilon_{3}e^{i\phi_3}+\delta) 
 & r_1 \\ 
 r_1(1-\epsilon_{1}e^{i\phi_1}-\epsilon_{3}e^{i\phi_3}+\delta)   & r_2^2e^{2i\gamma}(1-\epsilon_{2}e^{i\phi_2})  +1-2\epsilon_{3}e^{i\phi_3}+\delta & r_2^2e^{2i\gamma}(1-\epsilon_{2}e^{i\phi_2})  \\
 r_1  & r_2^2e^{2i\gamma}(1-\epsilon_{2}e^{i\phi_2}) &  r_2^2e^{2i\gamma}+1 \end{array}\right).\nonumber\\
\label{mtbmnu30}
\end{eqnarray}
\section{Connection to observables}
The broken $\mu\tau$ symmetric forms of $M_{\nu}^{\epsilon,\delta}$, 
cf. (\ref{mnuepd}), for the allowed three and four zero textures, 
viz. (\ref{mtbmnu40})
 and (\ref{mtbmnu30}), can be written in a unified compact form:
\begin{eqnarray}
 M_{\nu }^{\epsilon,\delta} &=& m\left[\left(\begin{array}{ccc}P&Q&Q\cr
                         Q &R& S\cr
                         Q&S&R\end{array}\right)\right.\nonumber\\&-&\left.\epsilon_1e^{i\phi_1}\left(\begin{array}{ccc}x_1&x_2&x_3\cr
                         x_2 &x_4& x_5\cr
                         x_3&x_5&x_6\end{array}\right)%\right. \nonumber\\ \left.
  -\epsilon_2e^{i\phi_2}\left(\begin{array}{ccc}y_1&y_2&y_3\cr
                         y_2 &y_4& y_5\cr
                         y_3&y_5&y_6\end{array}\right)-\epsilon_3e^{i\phi_3}\left(\begin{array}{ccc}z_1&z_2&z_3\cr
                         z_2 &z_4& z_5\cr
                         z_3&z_5&z_6\end{array}\right)-\delta\left(\begin{array}{ccc}t_1&t_2&t_3\cr
                         t_2 &t_4& t_5\cr
                         t_3&t_5&t_6\end{array}\right)\right].\nonumber\\
\label{bmnug}
\end{eqnarray}
The above equation contains all the six allowed $M_{\nu }^{\epsilon,\delta}$'s: four (two) from four (three)
zero textures. Here $m$ can be $m_A$, $m_B$ or $m_C$ depending on the category and as given in Table I. The real symmetry breaking parameters 
$\delta$, $\epsilon_i$, $\phi_i$, which were introduced in (\ref{mdmtb40}), (\ref{mtbmnu40}), (\ref{mdmtb30}) and  (\ref{mtbmnu30}), are
universal quantities. Both $\delta$ and $\epsilon_i$ are kept small in 
magnitude ($\le 0.15$) to ensure the validity of keeping only the lowest 
order perturbation. By construction, $\epsilon_3$ vanishes for four zero textures. In contrast, the (generally complex) quantities $P$, $Q$, $R$, $S$ as well as
$x_{1,..,6}$, $y_{1,..,6}$ and $z_{1,..,6}$ of (\ref{bmnug}) vary from category to category, though remaining unchanged within each category. On the other hand, the quantities
 $t_{1,..,6}$ are real and vary from texture to texture within each category. The expressions for all these in terms of the real positive parameters 
 $k_{1,2}$/$l_{1,2}$/$r_{1,2}$ and phase $\alpha/\beta/\gamma$ are listed in Table II. In the limit of vanishing $\delta$ and $\epsilon_{1,2,3}$, the $\mu\tau$
symmetric forms of $M_{\nu A}$, $M_{\nu B}$, $M_{\nu C}$ are restored in terms of $P$, $Q$, $R$ and $S$. 

%\begin{longtable*}
\begin{table*}[!ht]
\caption{\label{tabxtot} Expressions for quantities appearing in 
$M_\nu^{\epsilon,\delta}$ of (13). Parameters $z_{1-6}$, 
not needed for 
four zero textures since $\epsilon_3=0$, have been kept blank for the latter.} 
\begin{tabular}{|c|c|c|c|c|c|c|}
\hline
& \multicolumn{4}{c|}{Four zero} & \multicolumn{2}{c|}{Three zero}\\
\hline
Quanatity&Category $A1$&Category $A2$&Category $B1$ &Category $B2$&Category $C1$&Category $C2$\\
% \hline
% $m$ &$-c^2/M_2$&$-c^2/M_2$&$-c^2/M_2$&$-c^2/M_2$&$-d^2/M_2$&$-d^2/M_2$\\
\hline
$P$&$k_1^2e^{2i\alpha}+2k_2^2$&$k_1^2e^{2i\alpha}+2k_2^2$&$l_1^2$&$l_1^2$&$2r_1^2$&$2r_1^2$\\
\hline 
$Q$&$k_2$&$k_2$&$l_1l_2e^{i\beta}$&$l_1l_2e^{i\beta}$&$r_1$&$r_1$\\
\hline 
$R$&$1$&$1$&$l_2^2e^{2i\beta}+1$&$l_2^2e^{2i\beta}+1$&$r_2^2e^{2i\gamma}+1$&$r_2^2e^{2i\gamma}+1$\\
\hline 
$S$&$0$&$0$&$l_2^2e^{2i\beta}$&$l_2^2e^{2i\beta}$&$r_2^2e^{2i\gamma}$&$r_2^2e^{2i\gamma}$\\
\hline 
$x_1$&$2k_2^2$&$2k_2^2$&$0$&$0$&$2r_1^2$&$2r_1^2$\\
\hline 
$x_2$&$k_2$&$0$&$l_1l_2e^{i\beta}$&$l_1l_2e^{i\beta}$&$0$&$r_1$\\
\hline 
$x_3$&$0$&$k_2$&$0$&$0$&$r_1$&$0$\\
\hline 
$x_4$&$0$&$0$&$2l_2^2e^{2i\beta}$&$2l_2^2e^{2i\beta}$&$0$&$0$\\
\hline 
$x_5$&$0$&$0$&$l_2^2e^{2i\beta}$&$l_2^2e^{2i\beta}$&$0$&$0$\\
\hline 
$x_6$&$0$&$0$&$0$&$0$&$0$&$0$\\
\hline 
$y_1$&$0$&$0$&$0$&$0$&$0$&$0$\\
\hline 
$y_2$&$k_2$&$k_2$&$0$&$0$&$0$&$0$\\
\hline 
$y_3$&$0$&$0$&$0$&$0$&$0$&$0$\\
\hline 
$y_4$&$2$&$2$&$2$&$2$&$2r_2^2e^{2i\gamma}$&$2r_2^2e^{2i\gamma}$\\
\hline 
$y_5$&$0$&$0$&$0$&$0$&$r_2^2e^{2i\gamma}$&$r_2^2e^{2i\gamma}$\\
\hline 
$y_6$&$0$&$0$&$0$&$0$&$0$&$0$\\
\hline 
$z_1$&-&-&-&-&$0$&$0$\\
\hline 
$z_2$&-&-&-&-&$r_1$&$r_1$\\
\hline 
$z_3$&-&-&-&-&$0$&$0$\\
\hline 
$z_4$&-&-&-&-&$2$&$2$\\
\hline 
$z_5$&-&-&-&-&$0$&$0$\\
\hline 
$z_6$&-&-&-&-&$0$&$0$\\
\hline 
$t_1$&$-k_2^2$&$-k_2^2$&$0$&$0$&$-r_1^2$&$-r_1^2$\\
\hline
$t_2$&$-k_2$&$0$&$0$&$0$&$0$&$-r_1$\\
\hline
$t_3$&$0$&$-k_2$&$0$&$0$&$-r_1$&$0$\\
\hline
$t_4$&$-1$&$0$&$-1$&$0$&$0$&$-1$\\
\hline
$t_5$&$0$&$0$&$0$&$0$&$0$&$0$\\
\hline
$t_6$&$0$&$-1$&$0$&$-1$&$-1$&$0$\\
\hline
\end{tabular}
%\end{ruledtabular}
\end{table*}
%\end{longtable*}
%
In order to connect to observable quantities, we need
 to construct $H_{\nu}^{\epsilon,\delta}$, cf.(\ref{hepd}). 
Starting from (\ref{bmnug}), we obtain
\begin{eqnarray}
H_{\nu}^{\epsilon,\delta} &=& m^2\left[\left(\begin{array}{ccc}
|P|^2 +2|Q|^2& PQ^\star + Q(R^\star+S^\star)&  PQ^\star + Q(R^\star+S^\star)\cr
 P^\star Q + Q^\star(R+S)&|Q|^2+|R|^2+|S|^2&|Q|^2+RS^\star+R^\star S\cr
  P^\star Q + Q^\star(R+S)&|Q|^2+R^*S+RS^*&|Q|^2+|R|^2+|S|^2
 \end{array}\right)\right.\nonumber\\
 &-&\left.\epsilon_1\left(\begin{array}{ccc}u_1&u_2^*&u_3^*\cr
                         u_2 &u_4& u_5^*\cr
                         u_3&u_5&u_6\end{array}\right)-\epsilon_2\left(\begin{array}{ccc}v_1&v_2^*&v_3^*\cr
                         v_2 &v_4& v_5^*\cr
                         v_3&v_5&v_6\end{array}\right)-\epsilon_3\left(\begin{array}{ccc}w_1&w_2^*&w_3^*\cr
                         w_2 &w_4& w_5^*\cr
                         w_3&w_5&w_6\end{array}\right)-\delta\left(\begin{array}{ccc}s_1&s_2^*&s_3^*\cr
                         s_2 &s_4& s_5^*\cr
                         s_3&s_5&s_6\end{array}\right)\right].\nonumber\\
\label{bhg}
\end{eqnarray}
We have introduced in (\ref{bhg}) the quantities $u_i,~v_i,~w_i$ and $s_i$ ($i=1,..,6$) which are algebraic functions of  $P$, $Q$, $R$, $S$ and  $\phi_1$, $\phi_2$, $\phi_3$ as well as $x_i,~y_i,~z_i,~t_i$ ($i=1,..,6$). 
They appear explicitly
in Table II. Evidently, $u_k$, $v_k$, $w_k$, $s_k$ are real for $k=1,~4,~6$ and are generally complex otherwise.

The next point to note is this. With $i$ and $j$ running from $1$ to $6$, $u_i$ do not involve $y_j$, $z_j$, $t_j$, $\phi_2$ and $\phi_3$; similarly,
 $v_i$ do not involve $x_j$, $z_j$, $t_j$, $\phi_1$ and $\phi_3$; $w_i$ do not involve $x_j$, $y_j$, $t_j$, $\phi_1$ and $\phi_2$; 
 $s_i$ do not involve $x_j$, $y_j$, $z_j$, $\phi_1,~\phi_2$ and $\phi_3$. The explicit expressions for $u_i$, $v_i$, $w_i$ and $s_i$ are given in Table III.
It is interesting that they are related by certain substitution relations. Thus if $u_i$ are written as functions $f_i$ of set of appropriate variables, 
$v_i$ as well as $w_i$ and $s_i$ are the same functions $f_i$ of different sets of relevant variables.
\begin{eqnarray}
 u_i &=& f_i(x_1, x_2,...,x_6, \phi_1),\nonumber\\
 v_i &=& f_i(y_1, y_2,...,y_6, \phi_2),\nonumber\\
 w_i &=& f_i(z_1, z_2,...,z_6, \phi_3),\nonumber\\
 s_i &=&f_i(t_1, t_2,...,t_6, 0).
\label{uvwsf}
\end{eqnarray}
%\begin{longtable*}
\begin{table*}[!ht]
\caption{\label{exuvws} Expressions for $\mu\tau$ symmetry breaking quantities.} 
%\hline
\begin{center}
\begin{tabular}{|c|c|}
\hline
Functions & Expressions\\
\hline
$ u_1$ &$\left[P^*x_1+Q^*(x_2+x_3)\right]e^{i\phi_1}+c.c.$\\
\hline
$ u_2$ &$\left[P^*x_2+Q^*(x_4+x_5)\right]e^{i\phi_1}+\left[Qx_1^*+Rx_2^*+Sx_3^*\right]e^{-i\phi_1}$\\
\hline
$ u_3$ &$\left[P^*x_3+Q^*(x_5+x_6)\right]e^{i\phi_1}+\left[Qx_1^*+Sx_2^*+Rx_3^*\right]e^{-i\phi_1}$\\
\hline
$u_4$ &$\left[Q^*x_2+R^*x_4+S^*x_5\right]e^{i\phi_1}+c.c.$\\
\hline
$u_5 $&$\left[Q^*x_3+R^*x_5+S^*x_6\right]e^{i\phi_1}+\left[Qx_2^*+Sx_4^*+Rx_5^*\right]e^{-i\phi_1}$\\
\hline
$u_6$ &$\left[Q^*x_3+S^*x_5+R^*x_6\right]e^{i\phi_1}+c.c.$ \\
\hline
$ v_1$ &$\left[P^*y_1+Q^*(y_2+y_3)\right]e^{i\phi_2}+c.c.$\\
\hline
$ v_2$ &$\left[P^*y_2+Q^*(y_4+y_5)\right]e^{i\phi_2}+\left[Qy_1^*+Ry_2^*+Sy_3^*\right]e^{-i\phi_2}$\\
\hline
$ v_3$ &$\left[P^*y_3+Q^*(y_5+y_6)\right]e^{i\phi_2}+\left[Qy_1^*+Sy_2^*+Ry_3^*\right]e^{-i\phi_2}$\\
\hline
$v_4$ &$\left[Q^*y_2+R^*y_4+S^*y_5\right]e^{i\phi_2}+c.c.$\\
\hline
$v_5 $&$\left[Q^*y_3+R^*y_5+S^*y_6\right]e^{i\phi_2}+\left[Qy_2^*+Sy_4^*+Ry_5^*\right]e^{-i\phi_2}$\\
\hline
$v_6$ &$\left[Q^*y_3+S^*y_5+R^*y_6\right]e^{i\phi_2}+c.c.$ \\
\hline
$ w_1$ &$\left[P^*z_1+Q^*(z_2+z_3)\right]e^{i\phi_3}+c.c.$\\
\hline
$ w_2$ &$\left[P^*z_2+Q^*(z_4+z_5)\right]e^{i\phi_3}+\left[Qz_1^*+Rz_2^*+Sz_3^*\right]e^{-i\phi_3}$\\
\hline
$ w_3$ &$\left[P^*z_3+Q^*(z_5+z_6)\right]e^{i\phi_3}+\left[Qz_1^*+Sz_2^*+Rz_3^*\right]e^{-i\phi_3}$\\
\hline
$w_4$ &$\left[Q^*z_2+R^*z_4+S^*z_5\right]e^{i\phi_3}+c.c.$\\
\hline
$w_5 $&$\left[Q^*z_3+R^*z_5+S^*z_6\right]e^{i\phi_3}+\left[Qz_2^*+Sz_4^*+Rz_5^*\right]e^{-i\phi_3}$\\
\hline
$w_6$ &$\left[Q^*z_3+S^*z_5+R^*z_6\right]e^{i\phi_3}+c.c.$ \\
\hline
$ s_1$ &$\left[P^*t_1+Q^*(t_2+t_3)\right]+c.c.$\\
\hline
$ s_2$ &$\left[P^*t_2+Q^*(t_4+t_5)\right]+\left[Qt_1^*+Rt_2^*+Sx_3^*\right]$\\
\hline
$ s_3$ &$\left[P^*t_3+Q^*(t_5+t_6)\right]+\left[Qt_1^*+St_2^*+Rt_3^*\right]$\\
\hline
$s_4$ &$\left[Q^*t_2+R^*t_4+S^*t_5\right]+c.c.$\\
\hline
$s_5 $&$\left[Q^*t_3+R^*t_5+S^*t_6\right]+\left[Qt_2^*+St_4^*+Rt_5^*\right]$\\
\hline
$s_6$ &$\left[Q^*t_3+S^*t_5+R^*t_6\right]+c.c.$ \\
\hline
\end{tabular}
%\end{ruledtabular}
 \end{center}
\end{table*}
Before moving on to diagonalise $H_{\nu}^{\epsilon,\delta}$ of (\ref{bhg}), 
let us recall what happens in the $\mu\tau$ symmetric 
limit. In this case we can obtain 
\cite{Adhikary:2009kz}
 the neutrino mass squared differences and mixing angles 
in terms of three real functions $X_{1,2,3}$ of $P$, $Q$, $R$, $S$ with 
$X={(X_1^2+X_2^2)}^{1/2}$:
\begin{eqnarray}
 &&\Delta_{21}^2 \equiv m_2^2-m_1^2= m^2 X,\nonumber\\
 &&\Delta_{32}^2 \equiv m_3^2-m_2^2=\frac{m^2}{2}(X_3 - X),\nonumber\\
 &&\tan2\theta_{12} = \frac{X_1}{X_2},\nonumber\\
 &&m_{1,2} = {\left|\Delta_{21}^2\left(\frac{2-X_3\mp X}{2X}\right)\right|}^{1/2},\nonumber\\
 &&m_3  = \left|\Delta_{21}^2/X\right|^{1/2},
 \label{masangl} 
 \end{eqnarray}
with
\begin{eqnarray}
X_1 &=& 2\sqrt{2}|PQ^\star + Q(R^\star + S^\star)|,\nonumber\\
X_2 &=& |R+S|^2 - |P|^2,\nonumber\\
X_3 &=& |R + S|^2 - |P|^2-4(|Q|^2 +RS^\star +R^\star S).
 \label{x1x2x3} 
 \end{eqnarray}

Turning to $H_{\nu}^{\epsilon,\delta}$ of (\ref{bhg}) 
for broken $\mu\tau$ symmetry, it also can be diagonalised in a similar fashion to yield
$(\Delta_{21}^2)^{\epsilon,\delta}=(m_2^{\epsilon,\delta})^2-(m_1^{\epsilon,\delta})^2$,  
$(\Delta_{32}^2)^{\epsilon,\delta}=(m_3^{\epsilon,\delta})^2-(m_2^{\epsilon,\delta})^2$, $\theta_{12}^{\epsilon,\delta}$,
$\theta_{23}^{\epsilon,\delta}$, $\theta_{13}^{\epsilon,\delta}$. 
The superscripts explicitly signify that $\mu\tau$ symmetry breaking has been 
taken into account. Some details of the diagonalization procedure are 
given in the Appendix. The complicated algebraic expressions for those
quantities can be simplified by defining another set of functions $U_i$, $V_i$, $W_i$, $S_i$ ($i=1,..,6$) of the quantities introduced in (\ref{uvwsf}). 
 The detailed expressions are given in Table IV with 
$c_{12}=\cos\theta_{12},~s_{12}=\sin\theta_{12}$, $\theta_{12}$ 
being the unperturbed mixing angle
 between the first two generations and the phase $\psi$ being defined as
\begin{eqnarray}
\psi={\rm arg}~\left[P^*Q+Q^*(R+S)\right]. 
\end{eqnarray}
Note once again that each of the functions $U_i$, $V_i$, $W_i$, $S_i$ depends on the subset of the quantities $\left\{u_i,~v_i,~w_i,~s_i\right\}$ 
and is related to the others by a set of substitution rules analogous to (\ref{uvwsf}):
\begin{eqnarray}
 U_i &=& F_i(u_1,~ u_2,...,u_6),\nonumber\\
 V_i &=& F_i(v_1,~ v_2,...,v_6),\nonumber\\
 W_i &=& F_i(w_1,~ w_2,...,~w_6),\nonumber\\
 S_i &=&F_i(s_1,~ s_2,...,~s_6).
\label{UVWSF}
\end{eqnarray}
\begin{table}[!ht]
\caption{\label{exUVWS} Expressions for the functions appearing in (\ref{bmtres}).} 
%\hline
\begin{center}
\begin{tabular}{|c|c|}
\hline
Functions & Expressions\\
\hline
$U_1$ & $\frac{1}{2}[-2c_{12}^2u_1+\sqrt{2}c_{12}s_{12}\{(u_2+u_3)e^{-i\psi}+
(u_2^*+u_3^*)e^{i\psi}\}
-s_{12}^2(u_4+u_6+u_5+u_5^*)]$\\
\hline
$U_2 $&$\frac{1}{2}[-\sqrt{2}c_{12}^2(u_2+u_3)e^{-i\psi}+\sqrt{2}s_{12}^2(u_2^*+u_3^*)e^{i\psi}+
c_{12}s_{12}(u_4+u_6-2u_1+u_5+u_5^*)]$\\
\hline
$U_3$ & $\frac{1}{2}[\sqrt{2}c_{12}(u_2-u_3)e^{-i\psi}+s_{12}(u_6-u_4+u_5-u_5^*)]$\\
\hline
$U_4$ & $\frac{1}{2}[-2s_{12}^2u_1-\sqrt{2}c_{12}s_{12}\{(u_2+u_3)e^{-i\psi}+(u_2^*+u_3^*)e^{i\psi}\}
-c_{12}^2(u_4+u_6+u_5+u_5^*)]$\\
\hline
$U_5$ &$\frac{1}{2}[\sqrt{2}s_{12}(u_2-u_3)e^{-i\psi}-c_{12}(u_6-u_4+u_5-u_5^*)]$\\
\hline
$U_6$ &$\frac{1}{2}[u_5+u_5^*-u_4-u_6]$\\
\hline
$V_1$ & $\frac{1}{2}[-2c_{12}^2v_1+\sqrt{2}c_{12}s_{12}\{(v_2+v_3)e^{-i\psi}+
(v_2^*+v_3^*)e^{i\psi}\}
-s_{12}^2(v_4+v_6+v_5+v_5^*)]$\\
\hline
$V_2 $&$\frac{1}{2}[-\sqrt{2}c_{12}^2(v_2+v_3)e^{-i\psi}+\sqrt{2}s_{12}^2(v_2^*+v_3^*)e^{i\psi}+
c_{12}s_{12}(v_4+v_6-2v_1+v_5+v_5^*)]$\\
\hline
$V_3$ & $\frac{1}{2}[\sqrt{2}c_{12}(v_2-v_3)e^{-i\psi}+s_{12}(v_6-v_4+v_5-v_5^*)]$\\
\hline
$V_4$ & $\frac{1}{2}[-2s_{12}^2v_1-\sqrt{2}c_{12}s_{12}\{(v_2+v_3)e^{-i\psi}+(v_2^*+v_3^*)e^{i\psi}\}
-c_{12}^2(v_4+v_6+v_5+v_5^*)]$\\
\hline
$V_5$ &$\frac{1}{2}[\sqrt{2}s_{12}(v_2-v_3)e^{-i\psi}-c_{12}(v_6-v_4+v_5-v_5^*)]$\\
\hline
$V_6$ &$\frac{1}{2}[v_5+v_5^*-v_4-v_6]$\\
\hline
$W_1$ & $\frac{1}{2}[-2c_{12}^2w_1+\sqrt{2}c_{12}s_{12}\{(w_2+w_3)e^{-i\psi}+
(w_2^*+w_3^*)e^{i\psi}\}
-s_{12}^2(w_4+w_6+w_5+w_5^*)]$\\
\hline
$W_2 $&$\frac{1}{2}[-\sqrt{2}c_{12}^2(w_2+w_3)e^{-i\psi}+\sqrt{2}s_{12}^2(w_2^*+w_3^*)e^{i\psi}+
c_{12}s_{12}(w_4+w_6-2w_1+w_5+w_5^*)]$\\
\hline
$W_3$ & $\frac{1}{2}[\sqrt{2}c_{12}(w_2-w_3)e^{-i\psi}+s_{12}(w_6-w_4+w_5-w_5^*)]$\\
\hline
$W_4$ & $\frac{1}{2}[-2s_{12}^2w_1-\sqrt{2}c_{12}s_{12}\{(w_2+w_3)e^{-i\psi}+(w_2^*+w_3^*)e^{i\psi}\}
-c_{12}^2(w_4+w_6+w_5+w_5^*)]$\\
\hline
$W_5$ &$\frac{1}{2}[\sqrt{2}s_{12}(w_2-w_3)e^{-i\psi}-c_{12}(w_6-w_4+w_5-w_5^*)]$\\
\hline
$W_6$ &$\frac{1}{2}[w_5+w_5^*-w_4-w_6]$\\
\hline
$S_1$ & $\frac{1}{2}[-2c_{12}^2s_1+\sqrt{2}c_{12}s_{12}\{(s_2+s_3)e^{-i\psi}+
(s_2^*+s_3^*)e^{i\psi}\}
-s_{12}^2(s_4+s_6+s_5+s_5^*)]$\\
\hline
$S_2 $&$\frac{1}{2}[-\sqrt{2}c_{12}^2(s_2+s_3)e^{-i\psi}+\sqrt{2}s_{12}^2(s_2^*+s_3^*)e^{i\psi}+
c_{12}s_{12}(s_4+s_6-2s_1+s_5+s_5^*)]$\\
\hline
$S_3$ & $\frac{1}{2}[\sqrt{2}c_{12}(s_2-s_3)e^{-i\psi}+s_{12}(s_6-s_4+s_5-s_5^*)]$\\
\hline
$S_4$ & $\frac{1}{2}[-2s_{12}^2s_1-\sqrt{2}c_{12}s_{12}\{(s_2+s_3)e^{-i\psi}+(s_2^*+s_3^*)e^{i\psi}\}
-c_{12}^2(s_4+s_6+s_5+s_5^*)]$\\
\hline
$S_5$ &$\frac{1}{2}[\sqrt{2}s_{12}(s_2-s_3)e^{-i\psi}-c_{12}(s_6-s_4+s_5-s_5^*)]$\\
\hline
$S_6$ &$\frac{1}{2}[s_5+s_5^*-s_4-s_6]$\\
\hline
\end{tabular}
 \end{center}
\end{table}
The final expressions for the observable quantities in terms of their unperturbed values are
 \begin{eqnarray}
 (m_1^{\epsilon,\delta})^2 & = & m_1^2 + m^2\left[U_1\epsilon_1+V_1\epsilon_2+W_1\epsilon_3+S_1\delta\right],
 \nonumber\\
 (m_2^{\epsilon,\delta})^2 & = & m_2^2 + m^2\left[U_4\epsilon_1+V_4\epsilon_2+W_4\epsilon_3+S_4\delta\right],
 \nonumber\\
 (m_3^{\epsilon,\delta})^2 & = & m_3^2 + m^2\left[U_6\epsilon_1+V_6\epsilon_2+W_6\epsilon_3+S_6\delta\right].\nonumber\\
 (\Delta_{21}^2)^{\epsilon,\delta} & = & \Delta_{21}^2 + 
 m^2\left\{(U_4-U_1)\epsilon_1+(V_4-V_1)\epsilon_2+(W_4-W_1)\epsilon_3+(S_4-S_1)\delta
 \right\},\nonumber\\
 (\Delta_{32}^2)^{\epsilon,\delta} & = & \Delta_{32}^2 + m^2\left\{(S_6-S_4)\delta+
 (U_6-U_4)\epsilon_1+(V_6-V_4)\epsilon_2+(W_6-W_4)\epsilon_3\right\},\nonumber\\
 (\sin\theta_{12})^{\epsilon,\delta} & = & \left|s_{12} + c_{12}m^2\left\{\frac{S_2^*\delta+U_2^*\epsilon_1+
 V_2^*\epsilon_2+W_2^*\epsilon_3}{\Delta_{21}^2}\right\}\right|,\nonumber\\
 (\sin\theta_{23})^{\epsilon,\delta} & = & \left|\frac{1}{\sqrt 2}
 + \frac{s_{12}m^2}{\sqrt 2}\left\{\frac{S_3^*\delta+U_3^*\epsilon_1+
 V_3^*\epsilon_2+W_3^*\epsilon_3}{\Delta_{21}^2+\Delta_{32}^2}\right\}-\frac{c_{12}m^2}{\sqrt 2}
 \left\{\frac{S_5^*\delta+U_5^*\epsilon_1+
 V_5^*\epsilon_2+W_5^*\epsilon_3}{\Delta_{32}^2}\right\}\right|,\nonumber\\
 (\sin\theta_{13})^{\epsilon,\delta}& = &  
\left|c_{12}m^2\left\{\frac{S_3^*\delta+U_3^*\epsilon_1+
 V_3^*\epsilon_2+W_3^*\epsilon_3}{\Delta_{21}^2+\Delta_{32}^2}\right\}+s_{12}m^2\left\{\frac{S_5^*\delta +
 U_5^*\epsilon_1+
 V_5^*\epsilon_2+W_5^*\epsilon_3}{\Delta_{32}^2}\right\}\right|.\nonumber\\
\label{bmtres}
\end{eqnarray}
The CP-violating Jarlskog invariant, which is nonvanishing here because of $\mu\tau$ symmetry 
breaking, can be obtained from
\begin{eqnarray}
J_{\rm CP} &=& {\rm Im}\frac{(H_\nu^{\epsilon,\delta})_{12} (H_\nu^{\epsilon,\delta})_{23} (H_\nu^{\epsilon,\delta})_{31}}{
(\Delta_{21}^2)^{\epsilon,\delta} (\Delta_{32}^2)^{\epsilon,\delta} ({\Delta_{31}^2})^{\epsilon,\delta}}\nonumber\\
& =&\frac{1}{8}\sin2\theta_{12}^{\epsilon,\delta}\sin2\theta_{23}^{\epsilon,\delta}\sin2\theta_{13}^{\epsilon,\delta}\cos\theta_{13}^{\epsilon,\delta}\sin\delta_D^{\epsilon,\delta},
\label{CP}
\end{eqnarray}
where $\delta_D^{\epsilon,\delta}$  refres to Dirac phase in the PMNS matrix $U^{\epsilon,\delta}$.
\noindent
{\section{Numerical results and phenomenological discussion}}
\vskip 0.1in
\noindent
The breaking of $\mu\tau$ symmetry generates a nonzero $\theta_{13}$ as well as a deviation in $\theta_{23}$ from
 its maximal value of $\pi/4$. In the present work, 
the five experimentally measured inputs
\cite{Tortola:2012te,GonzalezGarcia:2012sz,Fogli:2012ua},
viz. the two mass
squared differences and the three mixing angles, are varied within 
their $3\sigma$ experimental ranges 
(Table V).
\begin{table}[!ht]
\caption{Input experimental values \cite{GonzalezGarcia:2012sz}} 
%\tbl{}
%\begin{tabular}{|c|c|} 
%\begin{center}
\begin{tabular}{|c|c|}
\hline
{ Quantity} & { $3\sigma$ ranges/other constraint}\\
\hline
$\Delta_{21}^2$ & $7.00<\Delta_{21}^2(10^{5}~ eV^{-2})<8.09$\\
$\Delta_{32}^2<0$ & $-2.649<\Delta_{32}^2(10^{3}~ eV^{-2})<-2.242$\\
%$\Delta_{32}^2>0$ & $2.276<\Delta_{32}^2(10^{3}~ eV^{-2})<2.695$\\
$\Delta_{32}^2>0$ & $2.195<\Delta_{32}^2(10^{3}~ eV^{-2})<2.625$\\
$\theta_{12}$ & $31.09^\circ<\theta_{12}<35.89^\circ$\\
$\theta_{23}$ & $35.80^\circ<\theta_{23}<54.80^\circ$\\
$\theta_{13}$ & $7.19^\circ<\theta_{13}<9.96^\circ$\\
$\sum_i m_i$ & $<0.5~eV$\\
$\delta_D$ & Unconstrained\\
\hline
\end{tabular}
%\end{center}
%\caption{Input experimental values \cite{Tortola:2012te}} 
\label{input}
\end{table}
We also constrain  the sum of the neutrino masses to be less than $0.5$ eV from cosmological considerations and 
leave the Dirac phase $\delta_D$ of the PMNS matrix $U$ to be unconstrained. 
On feeding the inputs, one can check
 which of the six textures can accommodate them. One universal remark can be made about the $\mu\tau$ symmetry 
breaking parameters $\epsilon_i,~\delta$. We restrict them to be less than $0.15$ in magnitude so that our
neglect of $\delta^2$, $\delta\epsilon_i$, $\epsilon_i\epsilon_j$ and higher order terms is numerically justified.
Thus $0<|\epsilon_i,\delta|<0.15$ in general.
There is nothing special about the number 0.15 which could have been 0.1. 
However, we want to maximize the ranges of $|\epsilon_i|$, $\delta$ though we 
neglect their second order contributions; this led to our choice of 0.15 as 
the upper limit. 
%%%%%%%%%%%%%%%%%%%%%%%%%%%%%%%%%%%%%%%%%%%%%%%
Among the six textures, we find only two of them, viz. $A1$ and $C1$, 
to survive the imposition of experimental data. The 
remaining four textures $A2$, $B1$, $B2$ and $C2$ are ruled out due to 
the following reasons.
\vskip 0.1in
$\bullet$ Texture A2 : ruled out due to a value of $\Delta_{32}^2$ outside 
the admissible range.
\vskip 0.1in
$\bullet$ Texture B1 : ruled out due to its $\theta_{13}$ value being 
outside the allowed interval.
\vskip 0.1in
$\bullet$ Texture B2 : excluded due to its  $\theta_{12}$ value being outside 
the observed range.
\vskip 0.1in
$\bullet$ Texture C2 : excluded owing to a $\theta_{13}$ value 
outside the allowed interval.
\vskip 0.1in
\noindent
%%%%%%%%%%%%%%%%%%%%%%%%%%%%%%%%%%%%%%%%%%%%%%%%%%%
We now make some phenomenological remarks on the surviving textures $A1$ 
and $C1$. Consider $A1$ first. The sensitivity of its three 
$\mu\tau$ symmetry breaking real parameters $\epsilon_1$, $\epsilon_2$ and 
$\delta$, cf.(8), to the observable quantities is discussed below.
\vskip 0.1in
\noindent
(i) The parameter that is least sensitive to the input data, in particular the 
value of $\theta_{13}$, is $\delta$. It is possible to accommodate all the 
data shown in Table I with any  $\delta$ within our chosen range. 
For simplicity, we set $\delta=0$ in our further analysis.
\vskip0.1in
\noindent
ii) Interestingly, it is $\epsilon_2$, which significantly affects 
the value of the angle $\theta_{23}$. 
For $\epsilon_2=0$ and very near zero but sizable values of 
$\epsilon_1$ and $\delta$, $\theta_{23}$ is always in the first 
octant, i.e. lower than $45^o$. 
On the other hand, a value of $\epsilon_2$ greater than 0.01 and 
in the range 0.01 - 0.15 takes to $\theta_{23}$ to the second octant, i.e. 
exceeding $45^o$. Thus the octant determination of $\theta_{23}$ will help 
pin down the value of $\epsilon_2$. 
\vskip 0.1in
\noindent
iii) The crucial role in generating a non zero $\theta_{13}$ is played by 
$\epsilon_1$. The range 
of $\epsilon_1$, that is needed to accommodate the observed $\theta_{13}$, 
is $0.05\leq\epsilon_1\leq 0.15$. 
The lower bound can be tinkered with a bit by suitably large values of 
$\epsilon_2$ and $\delta$, but it is a minor effect. 
\vskip 0.1in
\noindent
%%%%%%%%%%%%%%%%%%%%%%%%%%%%%%%%%%%%%%%%%%%%
\vskip 0.1in
Let us now turn to phases and masses. While one needs to restrict $\phi_1$
to $85^\circ\le\phi_1\le 100^\circ$ to fit the data, $\phi_2$ is found to 
be completely unrestricted. 
The allowed region of the $k_1$-$k_2$ parametric plane for the 
surviving texture is shown in the top left panel of  Fig.1. 
The permitted ranges of these parameters are found to be 
$1.06<k_1<1.42$, $0.25<k_2<0.70$. 
We find $\alpha$ to be constrained to be very close to $\pi/2$: 
$89^\circ<\alpha<90^\circ$. 
This $A1$ texture 
allows only an inverted ordering of the neutrino masses with 
$\Delta_{32}^2< 0$.
The allowed value of the sum $\sum_i m_i$ has been plotted against $m_1$ 
in the topmost right panel of Fig.1 and so have those of $m_2$ and $m_3$. 
We can say that $0.067\le m_1(eV)^{-1}\le 0.160$, 
$0.068\le m_2(eV)^{-1}\le 0.162$, 
$0.048\le m_3(eV)^{-1}\le 0.150$ and 
moreover $0.184\le (\sum_i m_i)(eV)^{-1}\le 0.480$. 
Thus a quasidegenerate 
neutrino mass spectrum with an inverted ordering is established. 
Going down, we have successively plotted 
the allowed magnitude of the Dirac phase $\delta_D$ vs that of the  
Jarlskog invariant $J_{CP}$ and the Majorana phases 
$\alpha_{M1}$ vs $\alpha_{M_2}$. 
We see that $0.001^\circ\le|\delta_D|\le 90^\circ$
 and $7.0\times{10^{-7}}\le |J_{CP}|\le 0.039$ whereas 
$29^\circ\le \alpha_{M1}\le 87^\circ$ and 
$12^\circ\le\alpha_{M2}\le 29^\circ$. 
\vskip 0.1in
\noindent
Coming to the three zero texture $C1$,  
we find in this case that 
there are four $\mu\tau$ symmetry breaking real parameters 
$\epsilon_1$, $\epsilon_2$, $\epsilon_3$ and $\delta$ as well as 
three phase angles 
$\phi_1$, $\phi_2$ and $\phi_3$. 
It is seen that with $\epsilon_1$ =  $\epsilon_2$
= 0 and nonzero values of the other parameters, it is 
possible to accommodate the input
experimental results. So, henceforth, we set $\epsilon_{1,2}=0$. 
One may also choose to set $\delta=0$; however, in that case, 
the range of 
$\theta_{13}$ comes out as a rather narrow one : 
$8.3^o<\theta_{13}<8.9^o$ and $\theta_{23}\sim 41.4^o$, i.e, the latter 
lies in the first octant. A different solution is found with 
$\theta_{23}\ge 45^o$ (i.e. in the second octant) for 
$\delta\ge 0.02$ 
and $\epsilon_3\ge 0.06$, but the lower bounds on those parameters have 
to be increased to 0.03 and 0.07 respectively in order to accommodate 
the full $3\sigma$ range of $\theta_{13}$. 
It may be noted that $\epsilon_3$ is the crucial parameter here for the 
generation of a nonzero $\theta_{13}$. On the other hand, the 
results are not 
sensitive to the upper bounds on $\delta$ and $\epsilon_3$. Finally, the 
effects of $\phi_i$ ($i=1,2,3$) are marginal; $\phi_{1,2,3}$ can swing from 
0 to $\pi$. 
%whereas $\phi_3$ is confined to $0\le \phi_3\le 150^o$. 
%\vskip 0.1in
%\noindent
This texture $C1$ also has $\Delta_{32}^2 < 0$. Plots, similar to those 
for texture $A1$, are shown in the top left panel of Fig.2 with the 
variation of parameters in the ranges $0.06\le \epsilon_3\le 0.15$ and 
$0.02\le \delta \le 0.15$. From these 
plots we obtain for the parameters $r_1$, $r_2$ and $\gamma$ of (12)
the ranges $0.655<r_1<1.130$, $0.968<r_2<1.350$,
, $89^\circ\le\gamma\le 90^\circ$. 
The top right panel of Fig.2, showing the neutrino mass interrelations, 
now implies 
$0.049\le m_1(eV)^{-1}\le 0.077$, $0.050\le m_2(eV)^{-1}\le 0.078$
 and $0.015\le m_3(eV)^{-1}\le 0.049$ and 
also $0.112\le (\sum_i m_i)(eV)^{-1}\le 0.203$. A weak inverted 
hierarchy is thus established. The bottom left plot of Fig.2 implies 
$0.003^\circ\le|\delta_D|\le 85^\circ$
 and $1.8\times{10}^{-6}\le |J_{CP}|\le 0.037$, while the bottom right plot leads to  
$-88^\circ\le\alpha_{M1}\le -25^\circ$ and
 $4^\circ\le\alpha_{M2}\le 46^\circ$. 
%%%%%%%%%%%%%%%%%%%%%%%%%%%%%%%%%%%%%%%%%%%%
\vskip 0.1in
\noindent
Finally, we should mention that we have performed similar numerical analyses
with the ranges of the input parameters obtained by the global analyses of 
Ref.\cite{Tortola:2012te} and Ref.\cite{Fogli:2012ua}. Though there are 
very minor variations in Figs.1 and 2, and in the consequent bounds given 
above, our basic conclusions are unaltered. 
In ending this section, let us comment on the feasibility of measuring 
the CP-violating parameters $\delta_D$ and 
$J_{CP}$ whose magnitudes we have plotted. 
Information on the latter can be extracted from experiments seeking  
CP violation with neutrino and antineutrino beams by 
measuring the difference in oscillation probabilities 
${\rm P}(\nu_\mu\rightarrow \nu_e)$ --
${\rm P}(\bar{\nu_\mu}\rightarrow {\bar{\nu_e}})$. This is 
reviewed in detail in Ref.\cite{Minakata:2008yz}. 
We need to state here 
that the 1$\sigma$ fit to $\delta_D$, reported in the global analysis 
of Ref.\cite{GonzalezGarcia:2012sz}, implies  
$\delta_D = {\left(300^{+66}_{-138}\right)}^o$.
The Majorana phases $\alpha_{M_1}$, $\alpha_{M_2}$ can be probed 
\cite{Bahcall:2004ip,Cremonesi:2012av} 
in neutrinoless double $\beta$ decay experiments but determining 
them is a challenging task. 
\vskip 0.25in
\noindent
\begin{figure}[!htb]
%\begin{flushleft}
%\includegraphics[width=8.0cm,height=8.0cm,angle=0]{four0-k1k2-garcia.png}
\includegraphics[width=3.8cm]{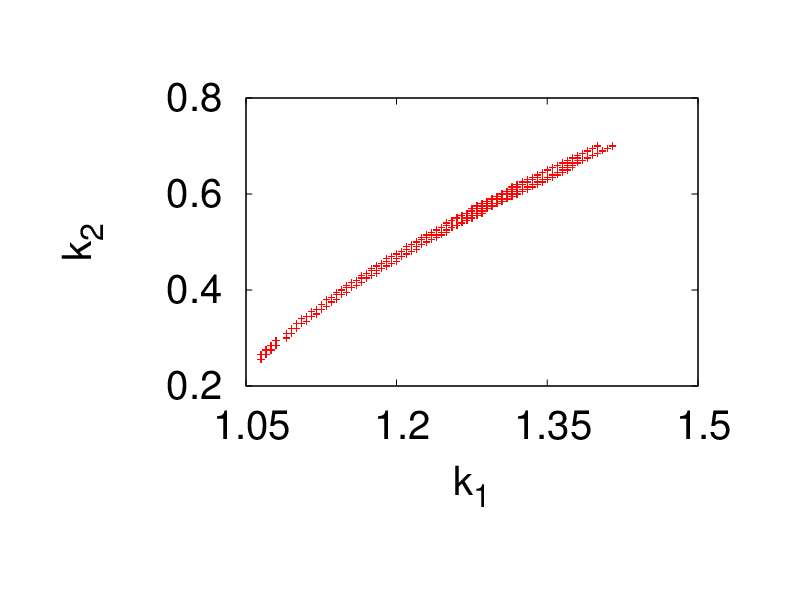}
\hskip 0.1in
\includegraphics[width=3.8cm]{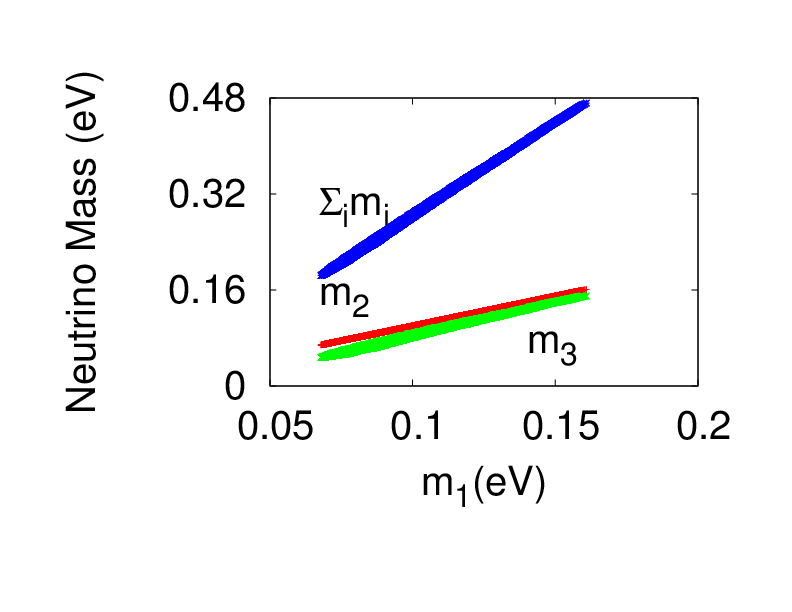}
\vskip 0.1in
\includegraphics[width=3.8cm]{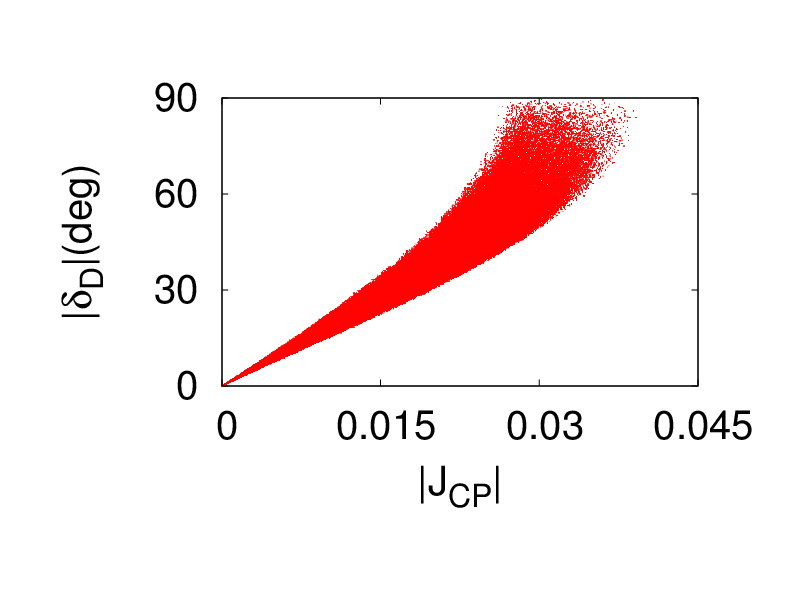}
\hskip 0.1in
\includegraphics[width=3.8cm]{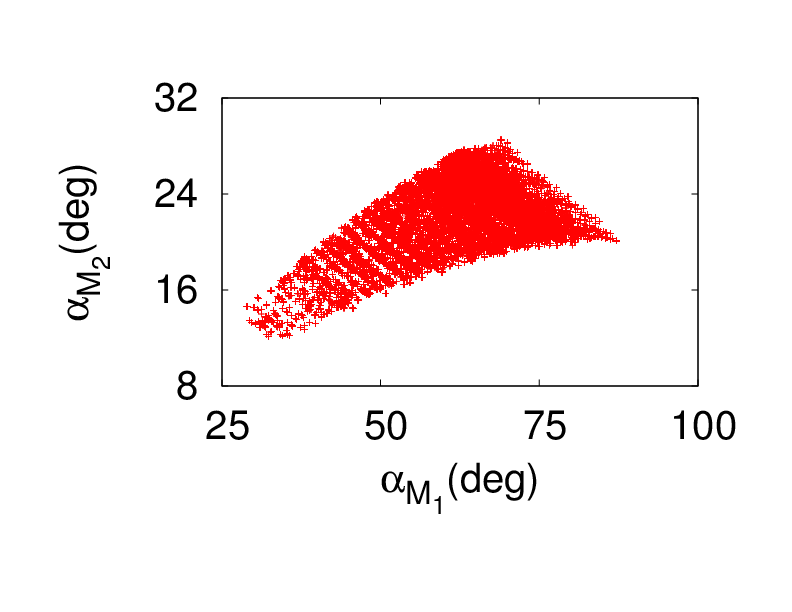}
%\end{flushleft}
\caption{\label{pl}(Color online) Panels showing allowed values: 
real parameters
$k_2$ vs $k_1$ and 
$\sum_i m_i$, $m_{2,3}$ vs $m_1$ (top) and
$|\delta_D|$ vs $|J_{CP}|$ and Majorana Phases $\alpha_{M1}$ vs
$\alpha_{M2}$ (bottom)
for category $A1$ of four zero textures.}
\end{figure}
%\newpage
%%%%%%%%%%%%%%%%%%   Figure 2  %%%%%%%%%%%%%%%%%%%%%%%%%%%%%%%
\begin{figure}[!htb]
%\begin{flushleft}
%\includegraphics[width=8.0cm,height=8.0cm,angle=0]{three0-r1r2-garcia.png}
\includegraphics[width=3.8cm]{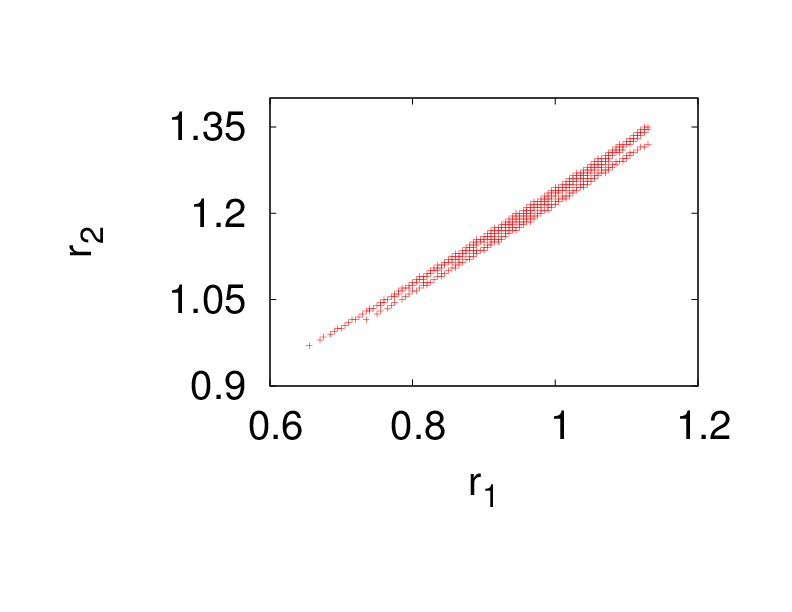}
\hskip 0.1in
\includegraphics[width=3.8cm]{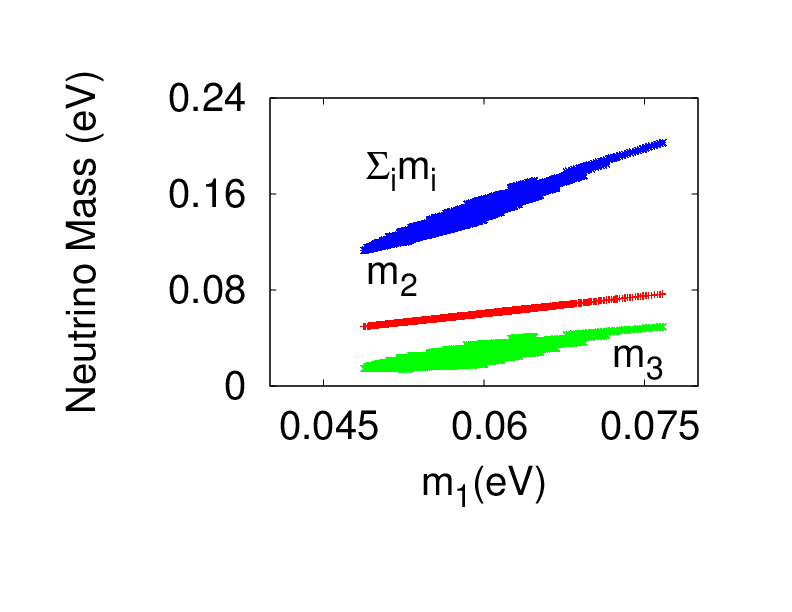}
\vskip 0.1in
\includegraphics[width=3.8cm]{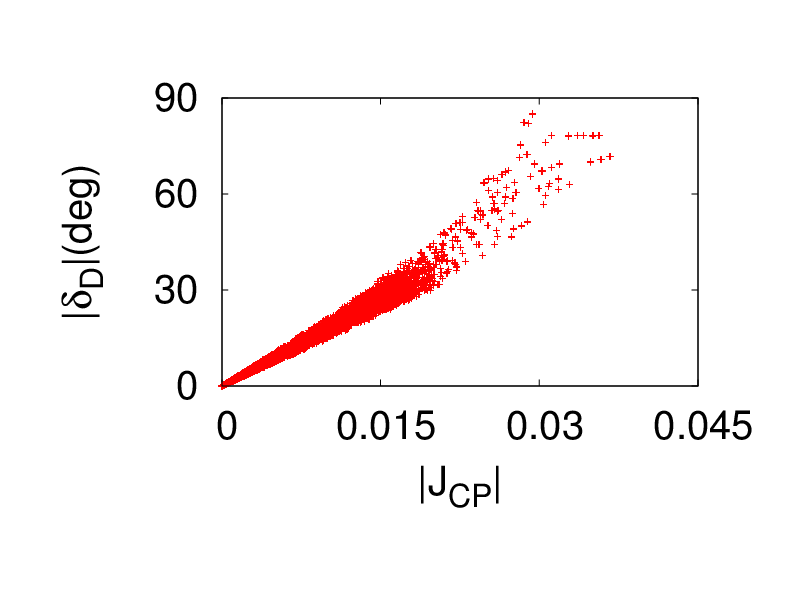}
\hskip 0.1in
\includegraphics[width=3.8cm]{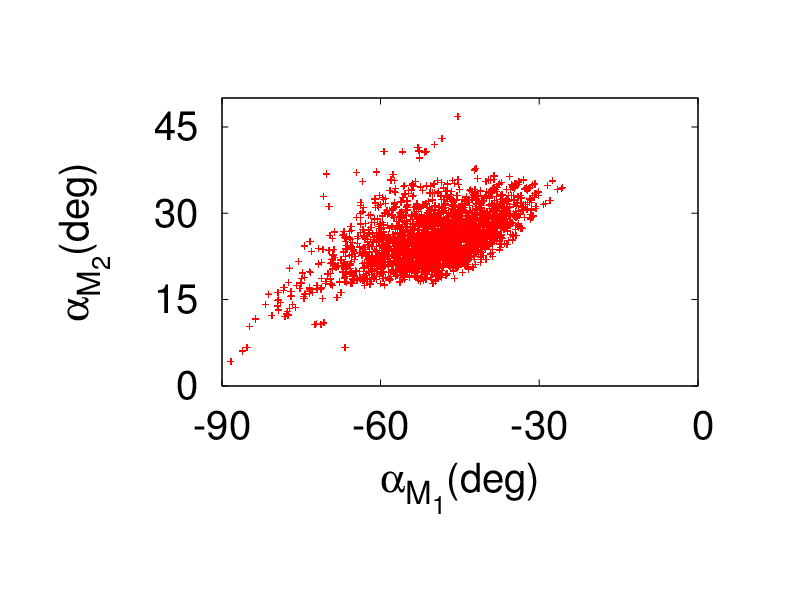}
% %\label{fig:marginalisedcontours}
%\end{flushleft} 
\caption{\label{pl}(Color online) Panels showing allowed values: real parameters
$r_2$ vs $r_1$  and $\sum_i m_i$, $m_{2,3}$ vs $m_1$ (top) and
$|\delta_D|$ vs $|J_{CP}|$ and Majorana Phases
$\alpha_{M1}$ vs $\alpha_{M2}$
(bottom)
for category $C1$ of
 three zero textures.}
\end{figure}
%%%%%%%%%%%%%%%%%%%%%%%%%%%%%%%%%%%%%
{\section {Summary and conclusion}}
In this paper we have considered $\mu\tau$ symmetric four and three
zero neutrino Yukawa textures allowed by our twin criteria of (1) no
massless and (2) no unmixed neutrino.
We have further introduced the most general $\mu\tau$ symmetry
breaking terms into these textures as a perturbation treated to the
lowest order, but keeping the textures intact. 
All these textures have then been subjected to the
minimal type-I seesaw in the weak basis defined by mass diagonal
(with real and positive values) charged leptons and heavy right 
chiral neutrinos. The resulting light neutrino Majorana mass matrix
$M_\nu$ has been constructed in each case and its consequences
compared quantitatively with the 3$\sigma$ ranges of five
experimental inputs: $\Delta_{21}^2$, $|\Delta_{32}^2|$, $\theta_{12}$,
$\theta_{23}$, $\theta_{13}$. It is found that, out of the originally 
allowed four 4-zero and two
3-zero textures, only one 4-zero and one 3-zero texture survive. The
survivor in each case can admit only an inverted mass ordering of the
light neutrinos. For the 4-zero case,  a quasidegenerate light
neutrino mass spectrum is established while the 3-zero case leads to a
weak inverted hierarchy. Allowed ranges of the neutrino masses, of
their Dirac and Majorana phases as well as their CP-violation strength
and of the magnitudes of the $\mu\tau$ symmetry breaking parameters have been 
 shown.
%%%%%%%%%%%%%%%%%%%%%%%%%%%%%%%%%Figure %%%%%%%%%%%%%%%%%%%
%\noindent
%\begin{figure}[!htb]
%\caption{\label{pl}(Color online) Panel showing allowed values: real parameters 
%$k_2$ vs $k_1$ and $r_2$ vs $r_1$ (top),
%$\sum_i m_i$ vs $m_1$ (second from top) and
%$|\delta_D|$ vs $|J_{CP}|$ (second from bottom) and Majorana Phases (bottom) 
%for category $A1$ (left) of four zero and category $C1$ (right) of
% three zero textures.}
%\begin{flushleft}
%\includegraphics[width=4.0cm,height=4.0cm,angle=0]{k1k2-40-replo.png}
%\hskip 0.1in
%\includegraphics[width=4.0cm,height=4.0cm,angle=0]{r1r2-30.png}
%\vskip 0.1cm
%\includegraphics[width=4.0cm,height=4.0cm,angle=0]{mass-40-new.png}
%\hskip 0.1in
%\includegraphics[width=4.0cm,height=4.0cm,angle=0]{mass-30-1.png}
%\vskip 0.1cm
%\includegraphics[width=4.0cm,height=4.0cm,angle=0]{jcp-delta-40.png}
%\hskip 0.1in
%\includegraphics[width=4.0cm,height=4.0cm,angle=0]{jcp-delta-3zero.png}
%\vskip 0.1in
%\includegraphics[width=4.0cm,height=4.0cm,angle=0]{majo-40.png}
%\hskip 0.1in
%\includegraphics[width=4.0cm,height=4.0cm,angle=0]{majophase-three0.png}
% %\label{fig:marginalisedcontours}
%\end{flushleft}
%\end{figure}
\newpage
\begin{center}
{\Large\bf Appendix}
\end{center}
\vskip 0.1in
\noindent
\appendix
\section{Diagonalization of $H_{\nu}^{\epsilon,\delta}$}
We describe here the methodolgy of diagonalizing the 
perturbed $H_{\nu}^{\epsilon,\delta}$ to obtain the results
given in (\ref{bmtres}). First of all, the unperturbed matrix $H_\nu$
can be 
diagonalized by the unitary matrix
\begin{eqnarray}
U=\left( \begin{array}{ccc}
e^{-i\psi} & 0 & 0\\
0&1&0\\
0&0&1          
         \end{array}\right)\left( \begin{array}{ccc}
c_{12} & s_{12} & 0\\
\frac{-s_{12}}{\sqrt 2}&\frac{c_{12}}{\sqrt 2}&\frac{-1}{\sqrt 2}\\
\frac{-s_{12}}{\sqrt 2}&\frac{c_{12}}{\sqrt 2}&\frac{1}{\sqrt 2}          
         \end{array}\right),
 \label{unu}
\end{eqnarray}
where $c_{12}=\cos{\theta_{12}}$, $s_{12}=\sin{\theta_{12}}$. 
Expressions for the unperturbed $\theta_{12}$ and  
neutrino masses have already been given in (\ref{masangl}). 
The phase $\psi$ is given in (18). In the first step of 
diagonalization, we rotate $H_{\nu}^{\epsilon,\delta}$ 
by the unperturbed $U$ of (\ref{unu}). 
In  consequence, off-diagonal terms  appear only as being linear 
in $\epsilon_i$, $\delta$, thereby vanishing in the unperturbed limit. 
The rotated form of
$H_{\nu}^{\epsilon,\delta}$ is
\begin{eqnarray}
U^\dagger H_{\nu}^{\epsilon,\delta}U&=& \left[\left(\begin{array}{ccc}
m_1^2&  0&0\\0&m_2^2&0\\0&0&m_3^2 \end{array}\right)
 +m^2\epsilon_1\left(\begin{array}{ccc}U_1&U_2^*&U_3^*\cr
                         U_2 &U_4& U_5^*\cr
                         U_3&U_5&U_6\end{array}\right)+m^2\epsilon_2\left(\begin{array}{ccc}V_1&V_2^*&V_3^*\cr
                         V_2 &V_4& V_5^*\cr
                         V_3&V_5&V_6\end{array}\right)\right.\nonumber\\&+&\left.m^2\epsilon_3\left(\begin{array}{ccc}W_1&W_2^*&W_3^*\cr
                         W_2 &W_4& W_5^*\cr
                         W_3&W_5&W_6\end{array}\right)+m^2\delta\left(\begin{array}{ccc}S_1&S_2^*&S_3^*\cr
                         S_2 &S_4& S_5^*\cr
                         S_3&S_5&S_6\end{array}\right)\right].
\end{eqnarray}
Looking at (A.2), we see the need to further rotate it by
 a matrix which deviates from the identity by 
terms linear in $\epsilon_i,~\delta$. For that purpose, let us consider 
the
following matrix
\newpage
\begin{eqnarray}
V^{\epsilon,\delta}=\left(\begin{array}{ccc} 1 &\epsilon_1 X_1^*+\epsilon_2X_2^*+\epsilon_3X_3^*
+\delta X_4^*& \epsilon_1 Y_1^*+\epsilon_2Y_2^*+\epsilon_3Y_3^*
+\delta Y_4^*\\-\epsilon_1 X_1-\epsilon_2X_2-\epsilon_3X_3-\delta X_4 &1& \epsilon_1 Z_1^*+\epsilon_2Z_2^*
+\epsilon_3Z_3^*
+\delta Z_4^*\\-\epsilon_1 X_1-\epsilon_2X_2-\epsilon_3X_3-\delta X_4 &-\epsilon_1 X_1-
\epsilon_2X_2-\epsilon_3X_3-\delta X_4&1\end{array}
\right).\nonumber\\
\end{eqnarray}
$V^{\epsilon,\delta}$ is unitary upto the neglect of terms beyond linear 
order in $\epsilon_i$ and $\delta$: 
${V^{\epsilon,\delta}}^\dagger V^{\epsilon,\delta}=I+O(\epsilon_i\epsilon_j)
+O(\epsilon_i\delta)+O(\delta^2)$. Thus we consider the final 
rotation 
\begin{equation}
U^{\epsilon,\delta} = U V^{\epsilon,\delta}
\end{equation}
and demand to have as a result a diagonal matrix with
squares of the perturbed masses as the entries. 
In this process we impose the vanishing condition on each coefficient 
of $\epsilon_i$ and $\delta$ in each off-diagonal element. 
That gives
us the expressions for $X_i$, $Y_i$, $Z_i$ ($i$=1-4) 
and hence the complete mixing matrix
\begin{eqnarray}
&&{U^{\epsilon,\delta}}^\dagger H_{\nu}^{\epsilon,\delta}U^{\epsilon,\delta}
=\left(\begin{array}{ccc} (m_1^{\epsilon,\delta})^2&  0&0\\0&(m_2^{\epsilon,\delta})^2&0\\
0&0&(m_3^{\epsilon,\delta})^2 \end{array}\right)\nonumber\\ 
&=& \left(\begin{array}{ccc}
\begin{array}{c}m_1^2+m^2U_1\epsilon_1+m^2V_1\epsilon_2\\+m^2W_1\epsilon_3 +
m^2S_1\delta \end{array}&  0&0\\0&\begin{array}{c}m_2^2+
m^2U_4\epsilon_1+m^2V_4\epsilon_2\\+m^2W_4\epsilon_3+m^2S_4\delta\end{array}&0\\0&0&\begin{array}{c}m_3^2 
+m^2U_6\epsilon_1+m^2V_6\epsilon_2\\+m^2W_6\epsilon_3+m^2S_6\delta\end{array}\end{array}\right).\nonumber\\
% +m^2\epsilon_1\left(\begin{array}{ccc}U_1&U_2^*&U_3^*\cr
%                          U_2 &U_4& U_5^*\cr
%                          U_3&U_5&U_6\end{array}\right)+m^2\epsilon_2\left(\begin{array}{ccc}V_1&V_2^*&V_3^*\cr
%                          V_2 &V_4& V_5^*\cr
%                          V_3&V_5&V_6\end{array}\right)\right.\nonumber\\&+&\left.m^2\epsilon_3\left(\begin{array}{ccc}W_1&W_2^*&W_3^*\cr
%                          W_2 &W_4& W_5^*\cr
%                          W_3&W_5&W_6\end{array}\right)+m^2\delta\left(\begin{array}{ccc}S_1&S_2^*&S_3^*\cr
%                          S_2 &S_4& S_5^*\cr
%                          S_3&S_5&S_6\end{array}\right)\right].
\label{fnr}
\end{eqnarray}
The vanishing condition for $(2,1)$ element after final rotation leads to 
 four equalities from the requirement of vanishing coefficients  
of $\epsilon_{1,2,3},~\delta$. Those are
\begin{eqnarray}
 X_1=\frac{m^2U_2}{m_2^2-m_1^2},\nonumber\\
X_2=\frac{m^2V_2}{m_2^2-m_1^2},\nonumber\\
X_3=\frac{m^2W_2}{m_2^2-m_1^2},\nonumber\\
X_4=\frac{m^2S_2}{m_2^2-m_1^2}.
\label{x}
\end{eqnarray}
Similarly, from the required vanishing of the 
$(3,1)$ element, we have
\begin{eqnarray}
Y_1=\frac{m^2U_3}{m_3^2-m_1^2},\nonumber\\
Y_2=\frac{m^2V_3}{m_3^2-m_1^2},\nonumber\\
Y_3=\frac{m^2W_3}{m_3^2-m_1^2},\nonumber\\
Y_4=\frac{m^2S_3}{m_3^2-m_1^2}.
\label{y}
\end{eqnarray}
The vanishing of the $(3,2)$ element yields  
\begin{eqnarray}
Z_1=\frac{m^2U_5}{m_3^2-m_2^2},\nonumber\\
Z_2=\frac{m^2V_5}{m_3^2-m_2^2},\nonumber\\
Z_3=\frac{m^2W_5}{m_3^2-m_2^2},\nonumber\\
Z_4=\frac{m^2S_5}{m_3^2-m_2^2}.
\label{z}
\end{eqnarray}
Finally, the mixing angles obtain, cf.(20), as 
\begin{eqnarray}
&& \sin{\theta_{12}^{\epsilon_i,\delta}}=|V_{12}|=|s_{12}+c_{12}\{ X_1^*\epsilon_1+X_2^*\epsilon_2
+X_3^*\epsilon_3+X_4^*\epsilon_4\}|,\nonumber\\
&&\sin{\theta_{13}^{\epsilon_i,\delta}}=|V_{13}|=|s_{12}\{ Z_1^*\epsilon_1+Z_2^*\epsilon_2
+Z_3^*\epsilon_3+Z_4^*\epsilon_4\}+c_{12}\{ Y_1^*\epsilon_1+Y_2^*\epsilon_2
+Y_3^*\epsilon_3+Y_4^*\epsilon_4\}|,\nonumber\\
&&\sin{\theta_{23}^{\epsilon_i,\delta}}=|V_{23}|=|\frac{1}{\sqrt 2} + \frac{s_{12}}{\sqrt 2}\{ Y_1^*\epsilon_1+Y_2^*\epsilon_2
+Y_3^*\epsilon_3+Y_4^*\epsilon_4\}
-\frac{c_{12}}{\sqrt 2}\{ Z_1^*\epsilon_1+Z_2^*\epsilon_2
+Z_3^*\epsilon_3+Z_4^*\epsilon_4\}|.\nonumber\\
\label{fnres}
\end{eqnarray}
Here we have substituted the expressions obtained 
for $X_i$, $Y_i$, $Z_i$ ($i$=1-4) from (\ref{x}), (\ref{y}) and (\ref{z})
directly into (\ref{fnres}). The perturbed neutrino squared masses in (20) 
follow from (A.5). 
% The research of P.R. has been supported in part by a DAE Raja Ramanna fellowship.  
% %%%%%%%%%%%%%%%%%%%%%%%%%%%%%%%%%%%%%%%%%%%%%%%%%%%%%
%\newpage

\end{document}